\documentclass[reprint, amsmath, amssymb, aps]{revtex4-2}
\usepackage{graphicx}
\usepackage{dcolumn}
\usepackage{bm}
\usepackage{siunitx}
\usepackage{amsmath}
\usepackage{xcolor}
\usepackage{soul}

\begin{document}

\title{Lower-depth programmable linear optical processors}

\author{Rui Tang}
 \email{ruitang@mosfet.t.u-tokyo.ac.jp}
\author{Ryota Tanomura}
\author{Takuo Tanemura}
\author{Yoshiaki Nakano}
\affiliation{Department of Electrical Engineering and Information Systems, The University of Tokyo, Tokyo 113-8656, Japan}
\date{\today}

\begin{abstract}
Programmable linear optical processors (LOPs) can have widespread applications in computing and information processing due to their capabilities to implement reconfigurable on-chip linear transformations. A conventional LOP that uses a mesh of Mach-Zehnder interferometers (MZIs) requires $2N+3$ stages of phase shifters for $N \times N$ matrices. However, it is beneficial to reduce the number of phase shifter stages to realize a more compact and lower-loss LOP, especially when long and lossy electro-optic phase shifters are used. In this work, we propose a novel structure for LOPs that can implement arbitrary matrices as long as they can be realized by previous MZI-based schemes. Through numerical analysis, we further show that the number of phase shifter stages in the proposed structure can be reduced to $N+2$ and $N+3$ for a large number of random dense matrices and sparse matrices, respectively. This work contributes to the realization of compact, low-loss, and energy-efficient programmable LOPs.
\end{abstract}

\maketitle

\section{\label{sec:level1}Introduction}
Programmable linear optical processors (LOPs) capable of implementing reconfigurable on-chip linear transformations have attracted increasing attention in recent years due to their promising applications in computing and information processing \cite{harris2018linear, bogaerts2020programmable, shen2017deep, perez2017multipurpose, dong2022high}. By programming the transfer matrix of optical processors, parallel tasks such as matrix multiplication can be directly performed in the optical domain, which may significantly reduce latency and lower energy consumption compared to their electronic counterparts \cite{nahmias2019photonic, zhou2022photonic, tang2022two, tait2022quantifying}.

So far, to implement an arbitrary matrix on a LOP, the matrix first needs to be decomposed into the product of two unitary matrices and a diagonal matrix via singular value decomposition \cite{miller2013self}, as illustrated in Fig. 1(a). The unitary matrices are realized using the universal multiport interferometer architectures which consist of a mesh of tunable Mach-Zehnder interferometers (MZIs) \cite{clements2016optimal, reck1994experimental, carolan2015universal, bell2021further, pai2019matrix}. Bell et al. have proposed a compact MZI-based structure that requires $N+2$ stages of phase shifters for realizing arbitrary $N \times N$ unitary matrices \cite{bell2021further}, as shown in Fig. 1(b). The diagonal matrix is realized using either a gain/absorber array or an MZI array. In practice, it is preferred to construct the entire LOP only using MZIs for easier fabrication and control. In this case, the structure of a LOP using the Bell structure is schematically shown in Fig. 1(c). Here, the last phase shifter stage in the section for $\rm \bf V$ and the first phase shifter stage in the section for $\rm \bf U$ have been absorbed into the MZI array for $\rm \bf \Sigma$. Therefore, for a conventional $N \times N$ LOP only using MZIs, in its most compact form, $2N+3$ phase shifter stages are required.

While thermo-optic (TO) phase shifters are widely used in existing devices \cite{shen2017deep, zhang2021optical}, a TO phase shifter typically consumes more than 1 mW power and is therefore not suitable for large-scale devices \cite{harris2014efficient, qiu2020energy, kita2021ultrafast}. To realize a high-speed and low-power LOP, electro-optic (EO) phase shifters are highly desirable, since a GHz modulation bandwidth and ultra-low power consumption during static operations can be easily achieved \cite{dong2010submilliwatt, han2017efficient, tang2022modulation, zheng2023electrooptically}. However, these phase shifters tend to have non-negligible insertion loss or can be millimeters in length, severely hindering their applications in LOPs. Therefore, it would be highly beneficial to reduce the number of phase shifter stages for the realization of compact and low-loss LOPs using EO phase shifters.

In this work, we propose a novel structure for programmable LOPs based on the concept of multiplane light conversion (MPLC) \cite{labroille2014efficient, fontaine2019laguerre, boucher2021full}. We show that using $N$ input/output ports out of an $N' \times N'$ universal multiport interferometer ($N' \geq 2N$), an arbitrary $N \times N$ matrix can be obtained as long as it can be realized by previous MZI-based schemes. Through numerical analysis, we further show that the number of phase shifter stages in the proposed structure can be reduced to $N+2$ and $N+3$ for a large number of random dense matrices and sparse matrices, respectively. This work contributes to the realization of compact, low-loss, and energy-efficient programmable LOPs.

\begin{figure*}[t]
\includegraphics[width=12cm]{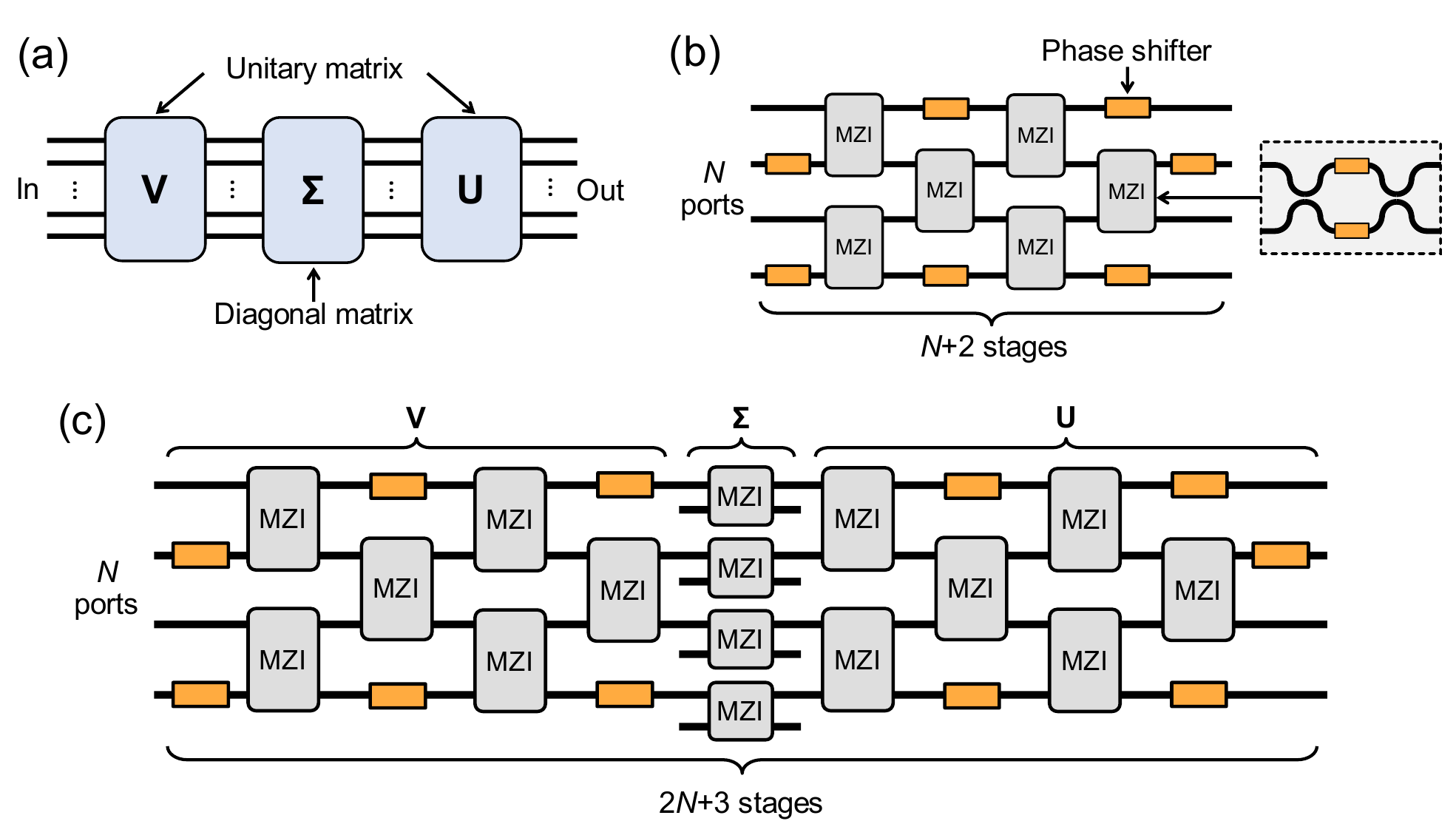}
\caption{\label{fig1} (a) Conventional programmable LOPs based on singular value decomposition. The target matrix is first decomposed into the product of two unitary matrices and a diagonal matrix. (b) A compact universal multiport interferometer proposed by Bell et al. for implementing arbitrary unitary matrices \cite{bell2021further}. $N+2$ stages of phase shifters are needed ($N=4$ in this figure). (c) The structure of a LOP using the Bell structure. The last phase shifter stage in the section for $\rm \bf V$ and the first phase shifter stage in the section for $\rm \bf U$ have been absorbed into the MZI array for $\rm \bf \Sigma$. $2N+3$ stages of phase shifters are needed ($N=4$ in this figure).}
\end{figure*}

\section{Principle}
\begin{figure}[b]
\includegraphics[width=8.6cm]{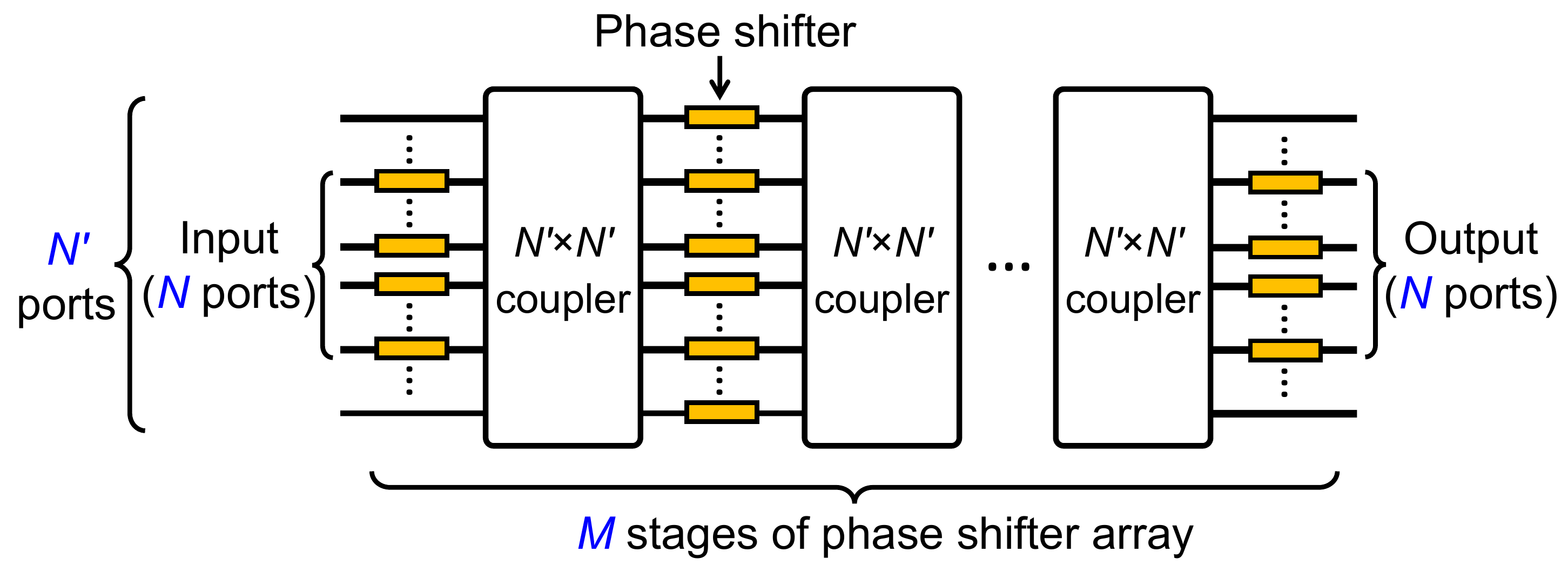}
\caption{\label{fig2} The proposed LOP structure. While the whole device has $N'$ ports, only $N$ ports are used as input/output ports. For unused input and output ports, phase shifters are not needed and thus are omitted in the first and last stages. For the stages in between two $N' \times N'$ couplers, all $N'$ phase shifters are used. The couplers can be multimode interference (MMI) couplers or multiport directional couplers. $M$ is the number of phase shifter arrays.}
\end{figure}

Our proposed structure is illustrated in Fig. \ref{fig2}. We use $N$ input/output ports out of an $N' \times N'$ universal multiport interferometer, which consists of cascaded phase shifter arrays and $N' \times N'$ couplers. The complex-valued transfer matrix can be calculated by multiplying the transfer matrices of all phase shifter arrays and couplers. The couplers can be multimode interference (MMI) couplers \cite{tang2017integrated, tang2018reconfigurable, tanomura2020monolithic}, or multiport directional couplers (MDCs) \cite{tang2017robust, tanomura2020robust, tang2021ten, kuzmin2021architecture}. As explained in the previous works \cite{tanomura2020robust, tang2021ten, tanomura2022scalable, saygin2020robust}, for schemes based on MPLC, the coupler is not required to have a specific transfer matrix as long as it properly scrambles the transmitted light.

For an $N \times N$ complex matrix $\rm \bf S$ that can be realized in previous schemes only using MZIs, as shown in Fig. 1(c), it can be written in the form of
\begin{equation}
\rm \bf S = U \Sigma V,
\label{eq1}
\end{equation}
where $\boldsymbol{{\rm U}}$ and $\boldsymbol{{\rm V}}$ are two $N \times N$ unitary matrices and $\boldsymbol{{\rm \Sigma}}$ is a $N \times N$ non-negative real diagonal matrix. Since we have assumed that $\boldsymbol{{\rm \Sigma}}$ is realized by an MZI array, it can be written in the form of
\begin{equation}
\boldsymbol{{\rm \Sigma}} =
    \begin{pmatrix}
    \sin\alpha_1 & & & \\
    & \sin\alpha_2 & & \\
    & & \ddots & \\
    & & & \sin\alpha_N \\
    \end{pmatrix},
\label{eq2}
\end{equation}
where $\alpha_i \in [0, \frac{\pi}{2}] (i = 1, 2, \cdots, N)$.

We now show that using $N$ input/output ports of a $N' \times N'$ universal multiport interferometer, which can implement arbitrary $N' \times N'$ unitary matrix, an arbitrary $\rm \bf S$ in the form of Eq. \ref{eq1} can be realized provided that $N' \geq 2N$. We first construct two $N' \times N'$ unitary matrices $\boldsymbol{{\rm U'}}$ and $\boldsymbol{{\rm V'}}$ from $\boldsymbol{{\rm U}}$ and $\boldsymbol{{\rm V}}$, respectively:
\begin{eqnarray}
\rm \bf U' =
    \begin{pmatrix}
        \rm \bf U & \\
        & \rm \bf O \\
    \end{pmatrix},
\label{eq3}
\\
\rm \bf V' =
    \begin{pmatrix}
        \rm \bf V & \\
        & \rm \bf P \\
    \end{pmatrix},
\label{eq4}
\end{eqnarray}
where $\rm \bf O$ and $\rm \bf P$ are arbitrary unitary matrices with $N'-N$ rows/columns. We then assume $N' \geq 2N$ and construct a $N' \times N'$ matrix $\rm \bf \Sigma'$ in the form of
\begin{equation}
\rm \bf \Sigma'=
\resizebox{0.85\hsize}{!}{
$\left(\begin{array}{cccc|cccc|c}
\sin\alpha_1 &  &  &  & j\cos\alpha_1 &  &  &  & \\
& \sin\alpha_2 &  &  &  & j\cos\alpha_2 &  &  &\\
&  & \ddots &  &  &  & \ddots &  & \\
&  &  & \sin\alpha_N &  &  &  & j\cos\alpha_N  &\\ \hline
j\cos\alpha_1 &  &  &  & \sin\alpha_1 &  &  &  &\\
& j\cos\alpha_2 &  &  &  & \sin\alpha_2 &  &  &\\
&  & \ddots &  &  &  & \ddots &  &\\
&  &  & j\cos\alpha_N &  &  &  & \sin\alpha_N &\\ \hline
&  &  &  &  &  &  &  & \rm \bf Q 
\end{array} \right),$}
\label{eq5}
\end{equation}
where $j$ is the imaginary unit and $\rm \bf Q$ is an arbitrary unitary matrix with $N'-2N$ rows/columns. $\rm \bf Q$ vanishes at the boundary case ($N' = 2N$). It is easy to verify that $\boldsymbol{{\rm \Sigma'}}$ is a unitary matrix and can be written as
\begin{equation}
\rm \bf \Sigma'=
    \begin{pmatrix}
        \rm \bf \Sigma & \rm \bf D \\
        \rm \bf D^\top & \rm \bf E
    \end{pmatrix},
\end{equation}
where
\begin{equation}
\rm \bf D =
    \begin{pmatrix}
    j\cos\alpha_1 & & & & 0 & \cdots \\
    & j\cos\alpha_2 & & & 0 & \cdots \\
    & & \ddots & & \vdots \\
    & & & j\cos\alpha_N & 0 & \cdots
    \end{pmatrix},
\end{equation}
\begin{equation}
\rm \bf E =
    \begin{pmatrix}
    \rm \bf \Sigma & \\
    & \rm \bf Q
    \end{pmatrix},
\end{equation}
and $\rm \bf D^\top$ is the transpose of $\rm \bf D$.

Now, we can see that the product of $\rm \bf U'$, $\rm \bf \Sigma'$, and $\rm \bf V'$ is
\begin{equation}
\begin{split}
\rm \bf S'=U'\Sigma'V'&= 
\left(\begin{array}{cc}
\rm \bf U & \\
& \rm \bf O
\end{array} \right)
\left(\begin{array}{cc}
\rm \bf \Sigma & \rm \bf D \\
\rm \bf D^\top & \rm \bf E 
\end{array} \right)
\left(\begin{array}{cc}
\rm \bf V & \\ 
& \rm \bf P
\end{array} \right)\\
&=\left(\begin{array}{cc}
\rm \bf S & \rm \bf UDP \\
\rm \bf OD^{\top}V & \rm \bf OEP
\end{array} \right),
\end{split}
\end{equation}
where $\rm \bf S$ is included as a submatrix. Since $\rm \bf U'$, $\rm \bf \Sigma'$, and $\rm \bf V'$ are all unitary matrices, $\rm \bf S'$ must also be a unitary matrix, which by definition can be realized by an $N' \times N'$ universal multiport interferometer. Because $\rm \bf S$ is a submatrix of $\rm \bf S'$, we reach the conclusion that using $N$ input/output ports of a $N' \times N'$ universal multiport interferometer, an arbitrary $\rm \bf S$ in the form of Eq. \ref{eq1} can be realized provided that $N' \geq 2N$. On the contrary, if $N' < 2N$, a unitary $\rm \bf \Sigma'$ can no longer be constructed and therefore $\rm \bf S'$ cannot be realized. It also can be verified that selecting $N$ ports from $N'$ ports can be arbitrary and does not have to be consecutive as in the above case, since permuting rows/columns in $\rm \bf S'$ does not affect its unitarity. Meanwhile, we note that other submatrices in $\rm \bf S'$ can be arbitrary since $\rm \bf O$, $\rm \bf P$, and $\rm \bf Q$ are all arbitrary. This suggests that there are redundant degrees of freedom in $\rm \bf S'$, and we may need fewer phase shifter stages to realize $\rm \bf S$. Considering the necessary degrees of freedom ($2N^2+N$), the stage number $M$ should be no less than $N+2$.

\section{Numerical analysis}
We use two different types of couplers: MMI couplers and MDCs, respectively, to investigate the necessary phase shifter stages in this scheme. For an ideal MMI coupler in which insertion loss and power imbalance are ignored, its transfer matrix is given in Ref. \cite{bachmann1994general}. For an MDC, we use the coupled mode theory to derive its transfer matrix \cite{huang1994coupled}. The details are provided in Appendix A. In real devices, the MMI coupler should be engineered to have a low insertion loss, which typically involves the use of waveguide tapers; the MDC should be designed to have a large coupling entropy \cite{tang2021ten}.
\begin{figure}[b]
\includegraphics[width=6cm]{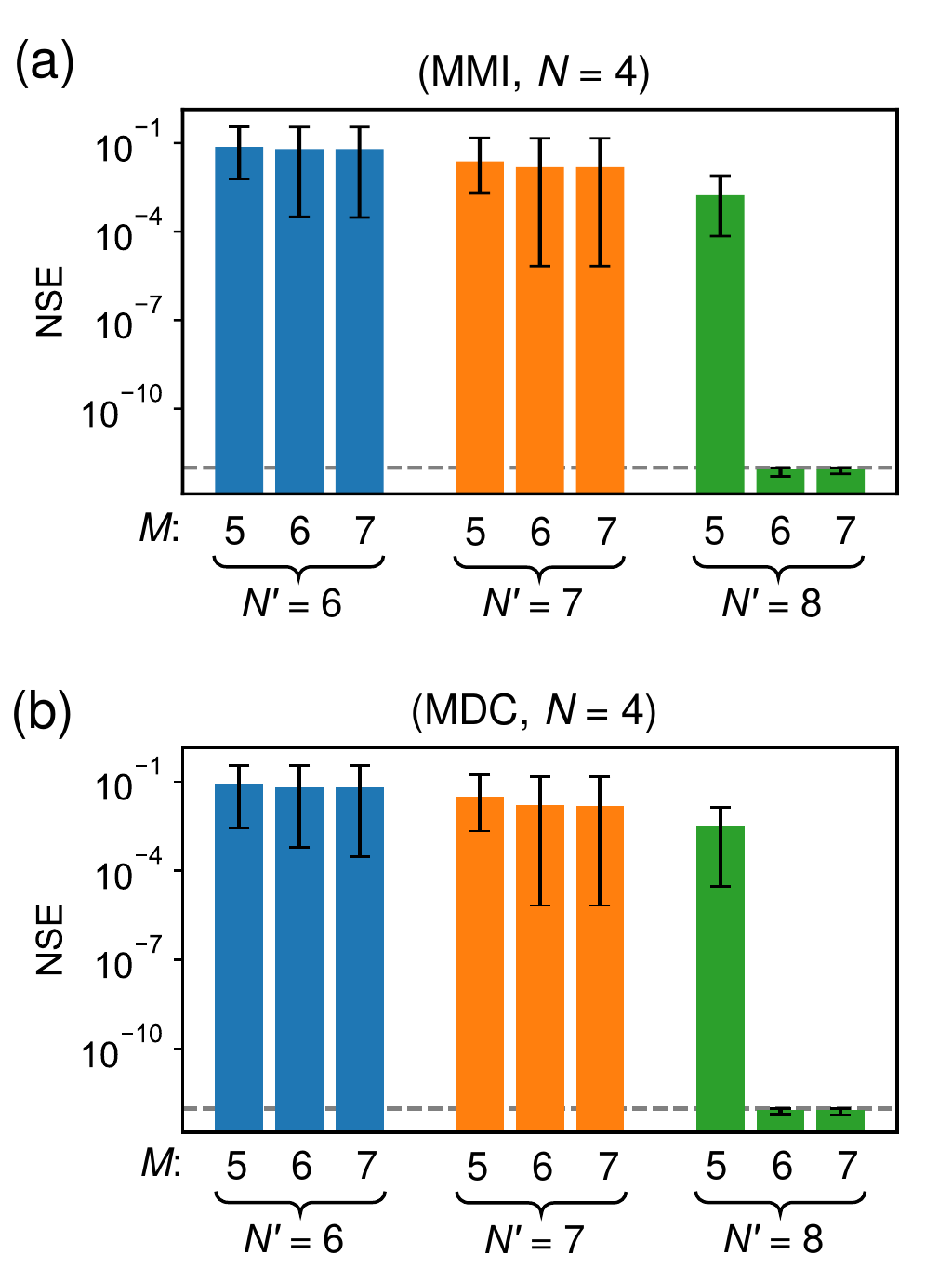}
\caption{\label{fig3} Average NSEs when the proposed LOPs are used to implement 100 random dense matrices with non-zero elements. The error bar represents the range between the maximum and minimum values among the 100 cases. The dash line indicates the NSE of $10^{-12}$. (a) LOPs using MMI couplers. (b) LOPs using MDCs.}
\end{figure}
\subsection{Dense matrices}
We consider the cases of various $N$, $N'$ and $M$. For each $N$, we generate 100 random $\rm \bf S$ as target matrices by multiplying randomly generated $\rm \bf U, \Sigma, V$, where $\rm \bf U$ and $\rm \bf V$ are Haar-random unitary matrices and the diagonal elements of $\rm \bf \Sigma$ are randomly sampled from the uniform distribution $[0, 1]$. In this way, all generated $\rm \bf S$ are dense matrices with non-zero elements. For each target matrix, we construct several LOPs with different $N'$ and $M$, and then optimize all the phase shifts to obtain the target matrix, using a covariance matrix adaptation evolution strategy (CMA-ES) algorithm \cite{hansen2023cma}. An open-source python package developed for the CMA-ES optimization is used \cite{cma-es}. For simplicity, the $N$ ports are chosen to be the middle ports of the $N'$ ports. After optimization, the difference between the target matrix $\rm \bf S$ and the obtained matrix $\boldsymbol{{\rm S}}^{\rm ob}$ is evaluated using the normalized squared error (NSE):
\begin{equation}
{\rm NSE} = \frac{1}{N} \sum\limits_{i=1}^{N}\sum\limits_{j=1}^{N} {\left|{\rm S}_{ij}-{\rm S}_{ij}^{\rm ob}\right|}^2.
\end{equation}
The NSE is also used as the cost function during optimizations, and the stopping criterion is set as either $\rm NSE < 10^{-12}$ or upon reaching a specific iteration number. For all optimizations, we set the initial phase shifts to $\pi$ and the update step size (parameter ``sigma" in the algorithm) to 2. In most cases, the algorithm converges successfully and returns the desired results.\\
\begin{figure}[t]
\includegraphics[width=6.5cm]{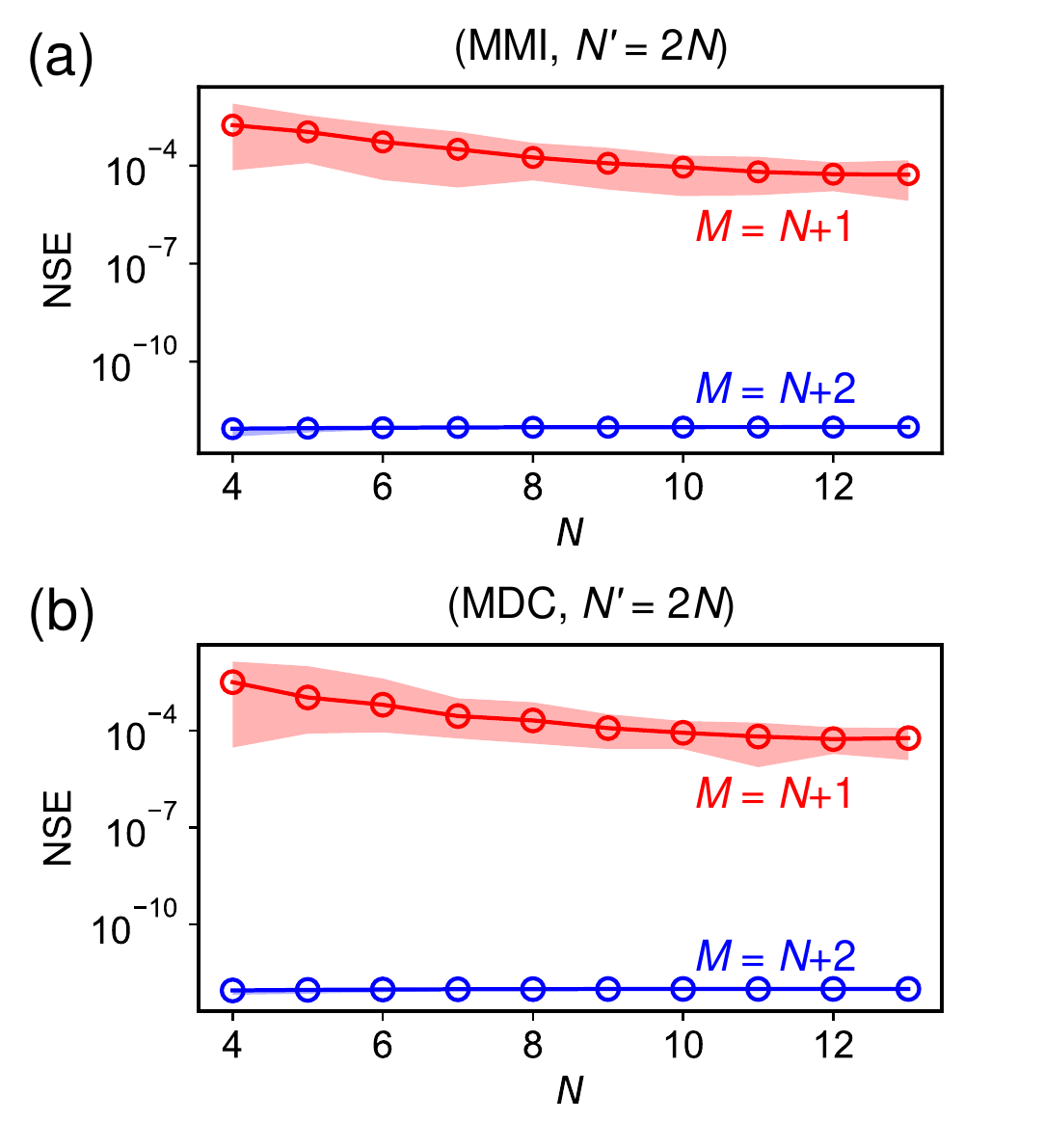}
\caption{\label{fig4} Average NSEs for random dense matrices with non-zero elements. The error band represents the range between the maximum and minimum values among the 100 cases. (a) LOPs using MMI couplers. (b) LOPs using MDCs.}
\end{figure}
Figure 3(a) and 3(b) show the average NSEs when the proposed LOPs are used to implement one hundred $4 \times 4$ random dense matrices ($N=4$), assuming the use of MMI couplers and MDCs, respectively. The error bar represents the range between the maximum and minimum value among the 100 cases. The dash line indicates the NSE of $10^{-12}$. We can see that very similar results are obtained for the two different couplers. As predicted in the previous section, an arbitrary matrix cannot be realized if $N'<2N$. Therefore, for $N'=6$ and $7$, the average NSE does not decrease significantly even when $M$ is increased. On the contrary, for $N'=8$, sufficiently small NSEs ($< 10^{-12}$) are obtained for all the 100 cases when $M \geq N+2$, indicating that all the desired matrices are almost perfectly realized. Figure 4 further shows the cases of various $N$. The error band represents the range between the maximum and minimum value among the 100 cases. For all desired matrices, sufficiently small NSEs are obtained for both types when $N'=2N$ and $M=N+2$. For $M=N+1$, although the NSE decreases slightly with increasing $N$, it cannot be sufficiently suppressed compared to the case of $M=N+2$. There is no notable difference between the LOP using MMI couplers and MDCs in terms of NSE. Therefore, while a rigorous proof is still lacking, the numerical analysis shows that $N+2$ stages of phase shifters are sufficient in this scheme for a large number of random dense matrices with non-zero elements.
\begin{figure}[t]
\includegraphics[width=6.5cm]{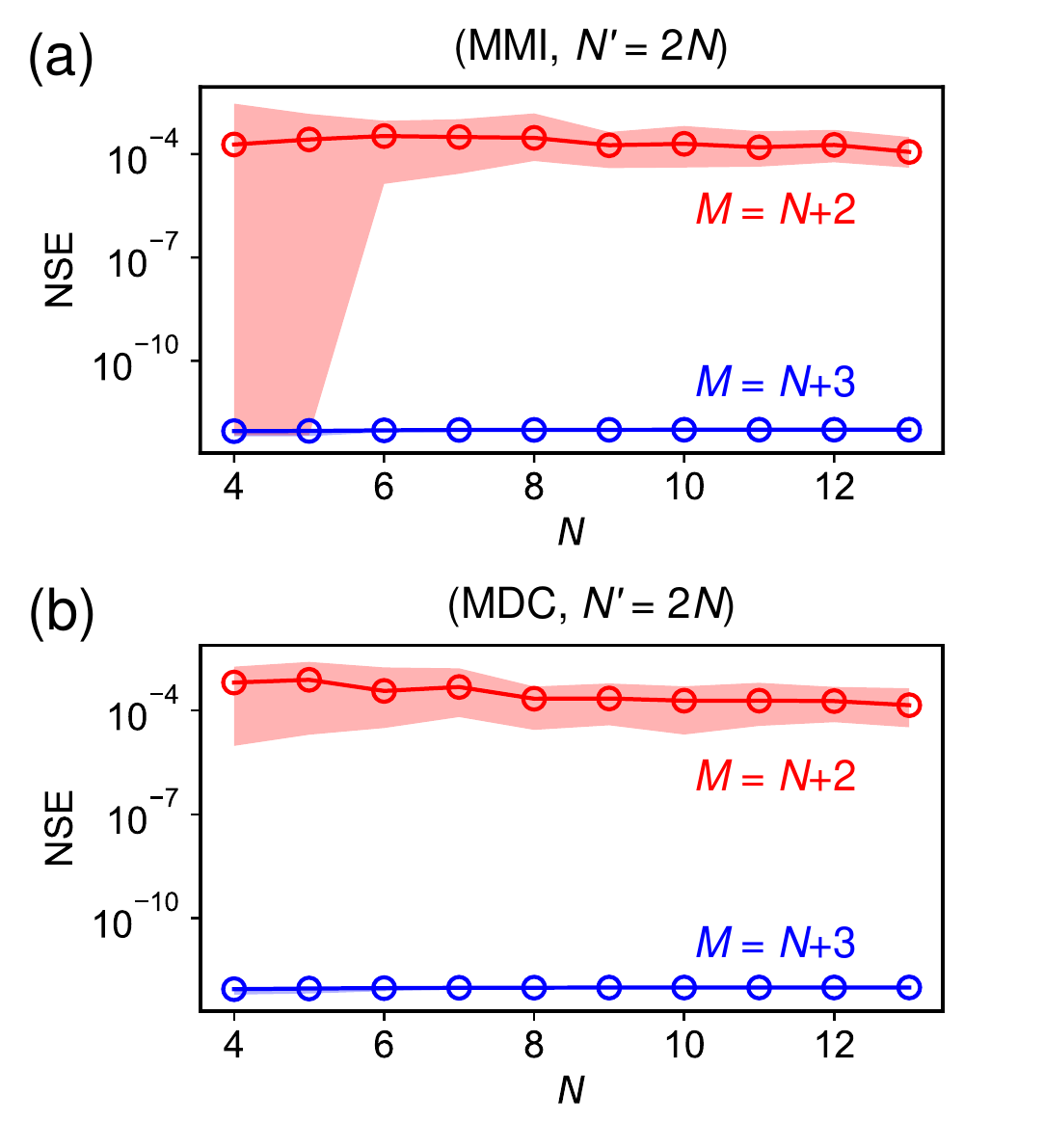}
\caption{\label{fig5} Average NSEs for random sparce matrices with one non-zero element. The error band represents the range between the maximum and minimum values among the 100 cases. (a) LOPs using MMI couplers. (b) LOPs using MDCs.}
\end{figure}
\subsection{Sparse matrices}
We then investigate the necessary stage number in this scheme for sparse matrices. Random dense matrices with non-zero elements are first generated following the same procedures described in Subsection A. Subsequently, sparse matrices with only one non-zero element are created by randomly setting elements to 0 in the dense matrices. We construct various LOPs to implement these sparse matrices. Figure 5 shows the average NSEs after optimizing the phase shifts. Each point again represents the average value of 100 random cases. It can be seen that all these sparse matrices are realized sufficiently well using $N+3$ stages. Although an extra stage is needed compared with the case of dense matrices, this increase is still acceptable from a practical standpoint. Additionally, we can note the differences in Fig. 5(a) and Fig. 5(b) for $N=4$ and $5$, which are attributed to the differences in the coupling entropy between the two couplers.
\subsection{Hardware-induced computational errors}
We further compare the hardware-induced computational error in this scheme and that in the conventional MZI-based scheme shown in Fig. 1(c). Computational errors can arise from many factors, such as fabrication errors and finite phase control resolutions. Since it is beyond the scope of this paper to investigate the effects of different error sources, here we focus solely on the phase quantization error, which is caused by the finite phase control resolution in practical devices. In the results shown in Figs. 3-5, all phase shifts have been assumed to have sufficiently fine resolutions. Now, we assume that the resolution of phase control is 10 bits and calculate the induced errors in both schemes. We use the same 100 random matrices in Fig. 4 as the targets and calculate the ideal phase configurations for each scheme, respectively. The phase configurations for the MZI-based scheme are derived using the decomposition algorithm proposed in Ref. 14. Then, these ideal phase configurations are replaced by 10-bit approximations, and the new matrices are calculated. Figure 6 shows the average NSEs between the target and obtained matrices. For this scheme, the results of using MMI couplers and using MDCs are almost the same. The MZI-based scheme has larger average NSEs than this scheme, and the difference increases with increasing $N$. This arises from the fact that an error occurring at an earlier stage affects all the following stages. The error accumulates along with the propagation of light in the circuit, and therefore, a deeper LOP has a larger error than a shallower one.

\begin{figure}[t]
\includegraphics[width=8cm]{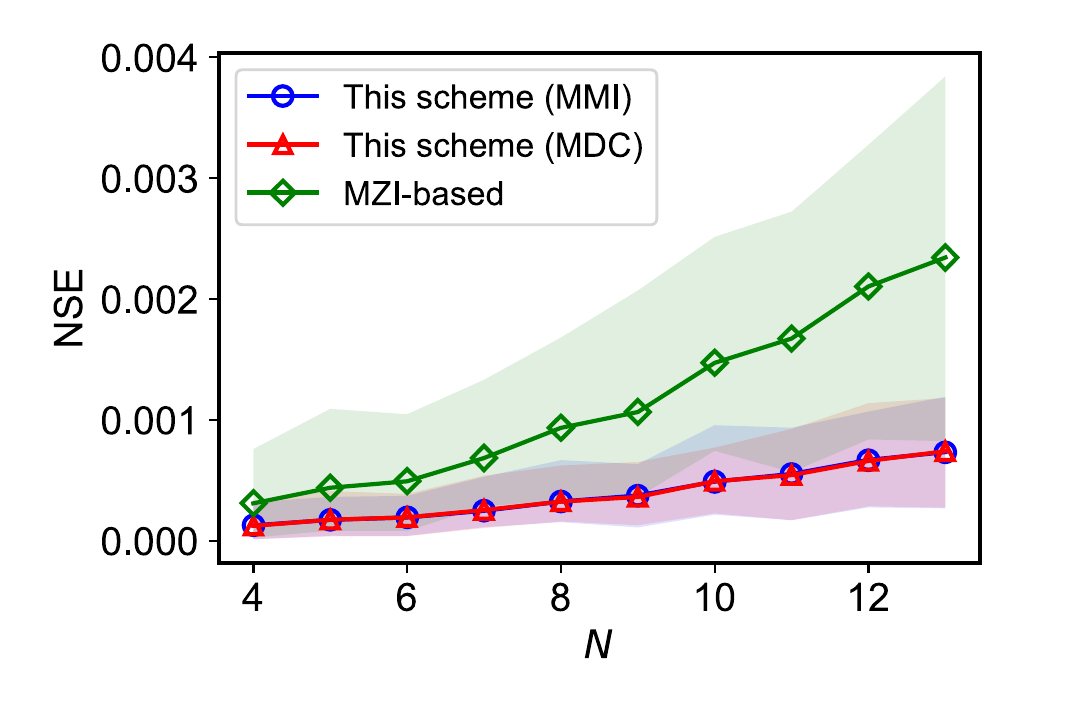}
\caption{\label{fig6} Average NSEs of this scheme ($N'=2N, M=N+2$) and the MZI-based scheme when all phase shifts are assumed to have 10-bit resolutions.}
\end{figure}

\section{Discussion}
It is natural to wonder if the proposed method can be applied into conventional schemes based on MZI meshes. Here, we show that although an arbitrary $\rm \bf S$ can be realized by increasing the port number to $2N$, the number of phase shifter stages can only be reduced by 1. We use $N=3$ and $N'=6$ as the example, as illustrated in Fig. 7. Without loss of generality, we assume that Ports 1-3 are used as the input/output ports. The transfer matrix of this $6 \times 6$ MZI mesh ${\rm \bf T}_{\rm mesh}$ can be calculated by sequentially multiplying the matrices of all components:
\begin{equation}
\begin{aligned}
    {\rm \bf T}_{\rm mesh} = {\rm \bf \Phi}_{4}^{8} {\rm \bf \Phi}_{2}^{8} {\rm \bf \Phi}_{6}^{7} {\rm \bf T}_{4,5}^{7} {\rm \bf T}_{2,3}^{7} {\rm \bf \Phi}_{1}^{7} {\rm \bf T}_{5,6}^{6} {\rm \bf T}_{3,4}^{6} {\rm \bf T}_{1,2}^{6} {\rm \bf \Phi}_{6}^{5} {\rm \bf T}_{4,5}^{5} {\rm \bf T}_{2,3}^{5}\\
    {\rm \bf \Phi}_{1}^{5} {\rm \bf T}_{5,6}^{4} {\rm \bf T}_{3,4}^{4} {\rm \bf T}_{1,2}^{4} {\rm \bf \Phi}_{6}^{3} {\rm \bf T}_{4,5}^{3} {\rm \bf T}_{2,3}^{3} {\rm \bf \Phi}_{1}^{3} {\rm \bf T}_{5,6}^{2} {\rm \bf T}_{3,4}^{2} {\rm \bf T}_{1,2}^{2} {\rm \bf \Phi}_{6}^{1} {\rm \bf \Phi}_{4}^{1} {\rm \bf \Phi}_{2}^{1} ,
\end{aligned}
\end{equation}
where ${\rm \bf T}_{i,i+1}^{k}$ corresponds to the MZI between the $i$-th and $(i+1)$-th ports on the $k$-th stage \cite{bell2021further}, ${\rm \bf \Phi}_{i}^{k}$ corresponds to the phase shifter outside MZIs on the $i$-th port, $k$-th stage and takes the form of
\begin{equation}
{\rm \bf \Phi}_{i}^{k} =
     \begin{pmatrix}
    1 & & & &  \\
    & \ddots & & & \\
    & & e^{j\cos\phi_i^k} & & \\
    & & & \ddots & \\
    & & & & 1
    \end{pmatrix},   
\end{equation}
where $\phi_i^k$ is the phase shift. For an arbitrary unitary matrix, each MZI matrix and phase shifter is determined recursively using the decomposition algorithm proposed by Bell et al \cite{bell2021further}. Since this MZI mesh can realize arbitrary $6 \times 6$ unitary matrices, according to the conclusion reached in Sec. II, an arbitrary $3 \times 3$ matrix can also be realized if it can be written in the form of Eq. 1. Meanwhile, it is known that multiplying a matrix with ${\rm \bf T}_{i,i+1}^{j}$ only affects the associated elements in the $i$-th and $(i+1)$-th rows/columns. Therefore, ${\rm \bf T}_{4,5}$, ${\rm \bf T}_{5,6}$ and ${\rm \bf \Phi}_{6}$ on all stages are redundant as they do not affect the matrix elements of our interest. It then becomes obvious that the required stages of phase shifters are $2N+2$.
\begin{figure}[b]
\includegraphics[width=8.5cm]{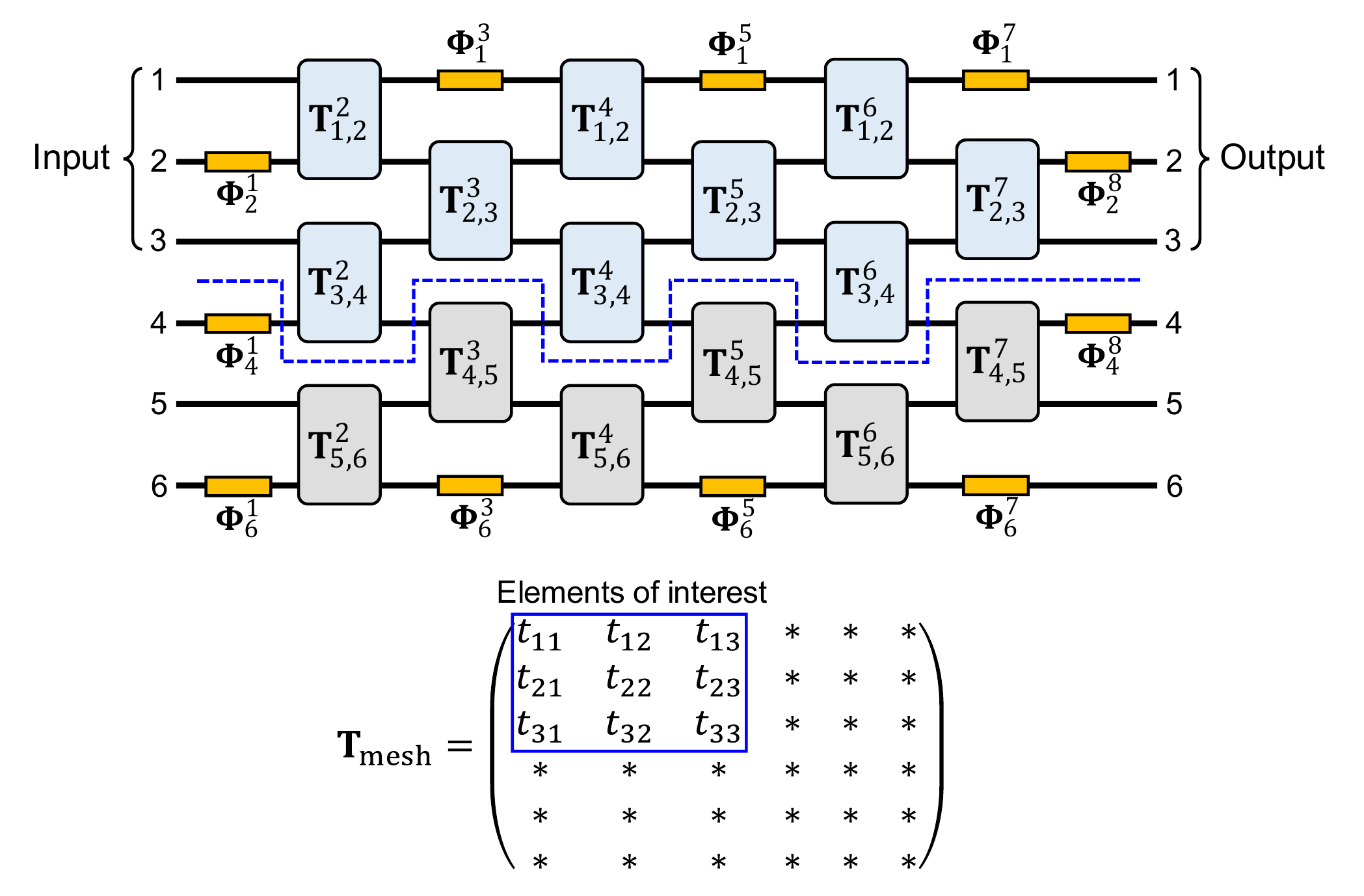}
\caption{\label{fig7} An example of using a $6 \times 6$ MZI-based universal multiport interferometer to implement $3 \times 3$ matrices. All MZIs and phase shifters below the dash line are redundant since they have no effect on the matrix elements of interest. $2N+2$ phase shifter stages are required ($N=3$ in this example).}
\end{figure}
The distinct difference in the required number of phase shifter stages between the MZI mesh and our proposed structure is attributed to the $N' \times N'$ couplers. The MMI coupler splits the input light from any input waveguide equally into all output waveguides. The MDC does not split the input light equally, but scrambles it in a way such that the light is not localized in only a few waveguides \cite{tanomura2020robust}. Therefore, the phase shifters following an $N' \times N'$ coupler change the matrix elements globally since each phase shifter affects the light from all ports in the previous stage. By contrast, a phase shifter in the MZI mesh changes the matrix elements locally since it only affects the light from two ports in the previous stage. This leads to the result that a large number of phase shifters in Fig. 6 are redundant.

Although the structure of the universal multiport interferometer is not new, the method of reducing the circuit depth by encoding a complex matrix as a block in a unitary matrix has not been proposed in previous works. While the number of ports has doubled in our scheme, the required number of phase shifter stages has been reduced from $2N+3$ in previous scheme to $N+2$ or $N+3$, which brings significant advantages in terms of insertion loss and footprint, especially when $N$ is large. In addition, the doubling of ports may not be a serious issue since there is no thermal crosstalk between the EO phase shifters, allowing them to be placed very close to each other. For unused output ports, optical power monitors can be integrated to provide real-time monitoring signals.

\section{conclusion}
We have proposed a novel structure for programmable LOPs based on the concept of MPLC. We have shown that using $N$ input/output ports out of an $N' \times N'$ universal multiport interferometer ($N' \geq 2N$), an arbitrary $N \times N$ matrix can be obtained as long as it can realized by conventional schemes based on MZI meshes. While a rigorous proof is yet to be provided, our numerical analysis suggests that the number of phase shifter stages in the proposed structure can be significantly reduced to $N+2$ for a large number of dense matrices, and $N+3$ for a large number of sparse matrices. We have further demonstrated that the same level of reduction cannot be achieved for the conventional scheme. This work contributes to the realization of compact, low-loss, and energy-efficient programmable LOPs.

\begin{acknowledgments}
This work was partly supported by JSPS KAKENHI Grant Numbers JP22K14298 and JP21K18168. R. Tang acknowledges fruitful discussions with Ryan Hamerly and Bryn Bell.
\end{acknowledgments}

\appendix

\section{Transfer matrix of the multiport directional coupler (MDC)}
We consider an MDC consisting of $N$ parallel straight waveguides with the same width and spacing. $N$ in this section should not be confused with the $N$ in the main context. We assume that the perturbation is weak so that the propagation constant in each waveguide is approximately the same as that in a single waveguide, and only coupling from nearest waveguides needs to be considered. Under these assumptions, the amplitudes of light in the MDC can be described by the coupled equations \cite{huang1994coupled}:
\begin{subequations}
\begin{eqnarray}
\frac{\mathrm{d} a_1}{\mathrm{d} z}&=&-j \beta a_1 - j \kappa_{1,2} a_2, \label{appa}
\\
\frac{\mathrm{d} a_i}{\mathrm{d} z}&=&-j \kappa_{i,i-1} a_{i-1} - j \beta a_i - j \kappa_{i,i+1} a_{i+1}, \label{appb}
\\
\frac{\mathrm{d} a_N}{\mathrm{d} z}&=&-j \kappa_{N,N-1} a_{N-1} - j \beta a_N , \label{appc}
\end{eqnarray}
\end{subequations}
where $z$ is the distance along the propagation direction, $a_i$ is the amplitude in the $i$-th waveguide ($1<i<N$), $\beta$ is the propagation constant, $\kappa_{i,i-1}$ is the coupling coefficient from the $(i-1)$-th to $i$-th waveguide. Since all the waveguides are assumed to have the same width and spacing, the coupling coefficient between each waveguide pair is the same and will hereafter be denoted as $\kappa$. The above equations can be written in the matrix form as:
\begin{equation}
\frac{\mathrm{d} \rm \bf A}{\mathrm{d} z} = -j\rm \bf HA,
\end{equation}
where
\begin{equation}
\rm \bf A=
    \begin{pmatrix}
    a_1 & a_2 & \cdots & a_N
    \end{pmatrix}^\top,
\end{equation}
\begin{equation}
\rm \bf H=
    \begin{pmatrix}
    \beta & \kappa & & & & \\
    \kappa & \beta & \kappa & & & \\
    & &  & \ddots & & \\
    & & & \kappa & \beta & \kappa \\
    & & & & \kappa & \beta 
    \end{pmatrix}.
\end{equation}
Since $\rm \bf H$ is a real Hermitian matrix, we can find a unitary matrix $\rm \bf B$ to diagonalize it into a diagonal matrix $\rm \bf C$:
\begin{equation}
\rm \bf C = B^{-1}HB.
\end{equation}
Then, by letting $\rm \bf W = B^{-1}A$ and substituting it into Eq. A2, we obtain:
\begin{equation}
\frac{\mathrm{d} \rm \bf W}{\mathrm{d} z} = -j\rm \bf CW.
\end{equation}
It is easy to see the solution of the above equation is:
\begin{equation}
\boldsymbol{{\rm W}} =
    \begin{pmatrix}
    e^{-j\mathrm{C_{11}}z} & e^{-j\mathrm{C_{22}}z} & \cdots & e^{-j\mathrm{C}_{N\!N}z}
    \end{pmatrix}^\top,
\end{equation}
where $\mathrm{C}_{ii}$ is the $i$-th diagonal element of $\rm \bf C$. It follows that
\begin{subequations}
\begin{eqnarray}
\boldsymbol{{\rm A}}(z) &=& \boldsymbol{{\rm B}}\begin{pmatrix}
    e^{-j\mathrm{C_{11}}z} & e^{-j\mathrm{C_{22}}z} & \cdots & e^{-j\mathrm{C}_{N\!N}z}
    \end{pmatrix}^\top,
\\
\boldsymbol{{\rm A}}(0) &=& \boldsymbol{{\rm B}}
    \begin{pmatrix}
    1 & 1 & \cdots & 1
    \end{pmatrix}^\top.
\end{eqnarray}
\end{subequations}
The above equations can be rewritten as:
\begin{equation}
\boldsymbol{{\rm A}}(z) = \boldsymbol{{\rm B}}
    \begin{pmatrix}
    e^{-j\mathrm{C_{11}}z} & & & \\
    & e^{-j\mathrm{C_{22}}z} & & \\
    & & \ddots & \\
    & & & e^{-j\mathrm{C}_{N\!N}z}     
    \end{pmatrix}
\boldsymbol{{\rm B^{-1}}}\boldsymbol{{\rm A}}(0),
\end{equation}
Therefore, for an MDC with a length of $L$, the transfer matrix is given by
\begin{equation}
\boldsymbol{{\rm T}} = \boldsymbol{{\rm B}}
    \begin{pmatrix}
    e^{-j\mathrm{C_{11}}L} & & & \\
    & e^{-j\mathrm{C_{22}}L} & & \\
    & & \ddots & \\
    & & & e^{-j\mathrm{C}_{N\!N}L}     
    \end{pmatrix}
\boldsymbol{{\rm B^{-1}}}.
\end{equation}

In this paper, we assume the use of silicon waveguides with core dimension of 500 $\times$ 220 \unit{\nm^2}. From numerical simulation, $\beta$ is found to be 9.91 \unit{\radian/\micro\meter} for transverse electric (TE) mode at 1550-\unit{nm} wavelength. For an $N$-port MDC, we assume $\kappa=0.05$ and select an $L$ so that small deviations in $\kappa$ and $L$ do not affect the overall performance of the LOP. Specifically, for $N=8, 10, 12, 14, 16, 18, 20, 22, 24, 26$, we choose $L$ to be 50, 60, 75, 85, 100, 120, 130, 140, 150, 160 \unit{\micro\meter}, respectively. Regardless of the value of $N$, two parameters: waveguide spacing and length, can be swept when designing an MDC. The design goal is to thoroughly scramble the light propagating through the MDC, which can be quantitatively described by the coupling entropy \cite{tang2021ten}. There exists a wide design region where the coupling entropy is large and does not change abruptly. The design parameters should be chosen from this region so that the device is robust to fabrication errors.

The above method yields unitary transfer matrices that simplify our analysis in this paper. However, if the distance between adjacent waveguides is small enough that the assumption of weak perturbation no longer holds, numerical approaches should be used to obtain more precise results \cite{cooper2009numerically}. 

\nocite{*}

\bibliography{references}

\end{document}